\documentclass[a4paper]{article}
\usepackage{amsmath}
\usepackage{amsfonts}
\usepackage{float}
\usepackage{bbm}
\usepackage{bm}
\usepackage{listings}
\usepackage{graphicx}
\usepackage{color}

\usepackage[linktocpage,hidelinks]{hyperref}
\usepackage{setspace}

\usepackage{caption}
\usepackage{subcaption}
\usepackage{ulem}
\usepackage{algorithm}
\usepackage[noend]{algpseudocode}

\usepackage{natbib}

\makeatletter
\def\BState{\State\hskip-\ALG@thistlm}
\makeatother

\DeclareMathOperator*{\Ex}{E}
\DeclareMathOperator*{\Var}{Var}
\DeclareMathOperator*{\Cov}{Cov}

\newcommand{\MVAR}{\text{MVAR}}

\DeclareMathOperator*{\VaR}{VaR}

\newcommand{\EFF}{\text{EFF}}
\newcommand{\MVP}{\text{MVP}}

\providecommand{\keyword}[1]{\textbf{\textit{Keywords: }} #1}
\providecommand{\corresp}[1]{\textbf{\textit{Correspondence: }} #1}
\providecommand{\email}[1]{\textbf{\textit{E-mail: }} #1}
\providecommand{\declar}[1]{\textbf{\textit{Declarations of interest: }} #1}

\title{Portfolio optimisation with mixture vector autoregressive models}

\author{Davide Ravagli \\ \href{davide.ravagli@manchester.ac.uk}{\normalsize davide.ravagli@manchester.ac.uk} 
	\and Georgi N. Boshnakov \\ \href{georgi.boshnakov@manchester.ac.uk}{\normalsize georgi.boshnakov@manchester.ac.uk} }

\date{%
	  Department of Mathematics \\%
      The University of Manchester \\%
      Manchester, United Kingdom\\[2ex]%
      }
                                        
\begin{document}

\maketitle


\corresp{Georgi Boshnakov \\ Department of Mathematics, The University of Manchester, \\
	Alan Turing Building, M13 9PY, Manchester, United Kingdom} \\
\email{georgi.boshnakov@manchester.ac.uk}

\begin{abstract}
Obtaining reliable estimates of conditional covariance matrices is 
an important task of heteroskedastic multivariate time series.
In portfolio optimisation and financial risk 
management, it is crucial to provide measures of uncertainty and risk
as accurately as possible.
We propose using mixture vector autoregressive (MVAR) models for portfolio
optimisation. Combining a mixture of distributions that depend on the recent
history of the process, MVAR models can accommodate asymmetry, multimodality,
heteroskedasticity and cross-correlation in multivariate time series data.
For mixtures of Normal components, we exploit a property of the multivariate
Normal distribution to obtain explicit formulas of conditional predictive
distributions of returns on a portfolio of assets. 
After showing how the method works, we perform a comparison with other relevant
multivariate time series models 
on real stock return data.  
\end{abstract}

\keyword{Forecasting;
	     Heteroskedasticity;
	     Mixture vector autoregressive model;
         MVAR model;
         Portfolio.
}

\declar{none.}

\section{Introduction}

Financial and econometric data often presents the feature of
heteroskedasticity. For multivariate time series, this implies that
the covariance matrix of an observation at a given time point depends 
upon the recent history of the process. This may be due to changes in 
the volatility of a single series, as well as in the 
cross-correlations between any two series of interest.
As a result, one cannot trust sample
estimates of the (unconditional) covariance matrix, or 
linear time series models 
to build reliable predictions about the future.
Therefore, obtaining reliable estimates of covariance matrices remains an important
challenge 
in portfolio optimisation and financial risk management which use, for instance,
modern portfolio theory \citep{Markowitz52}.
\citet{bollersev1988} and 
\citet{englekroner1995} pioneered in the attempt to model conditional covariance
matrices of predictors for multivariate time series with multivariate GARCH models,
using different parametrisations known respectively as VEC and 
BEKK. \citet{Engle2002} extended the idea of multivariate GARCH to
the so called Dynamic Conditional Correlation models, 
in which each element of the time-dependent covariance matrix of the data is 
modelled to follow a GARCH process. Such models have computational advantages over
multivariate GARCH models in that the number of
parameters to be estimated in the correlation process is independent of the number
of series to be correlated, by use of common parameters across
all correlations to be estimated.

Since then, much work has been done to develop multivariate GARCH models for
portfolio optimisation. In particular, many attempts have been made
in combining GARCH and factor models, with the aim of dimensionality reduction when
modelling large portfolios or panel data.
These models rely on the assumption that financial returns are described by
a small number of underlying common variables, or factors, 
which can be used to model
the data more parsimoniously. Although all equal in concept, different approaches
used different assumptions on such factors, and different techniques are 
used to derive them. For instance, \citet{alexander2000primer} uses a principal
components analysis in which factors are assumed to follow independent GARCH
 processes, 
whereas \citet{van2002go} considers the case in which factors are not orthogonal.
Finally, \citet{SANTOS2014606} introduced the dynamic factor GARCH model with
 time-varying factor loadings.
 
We propose using a mixture vector autoregressive (MVAR) model \citep{Fong2007} for
portfolio optimisation.
MVAR models are the multivariate extension 
of the mixture autoregressive (MAR) model by \citet{WongLi2000}. Combining
predictive distributions which depend on the recent history of the process,
MVAR models can accommodate asymmetry, multimodality, heteroskedasticity and 
cross-correlation in multivariate time series data. 
Theoretical properties of MVAR
were explored for the case of a multivariate Gaussian mixture in \citet{Fong2007}
and \citet{KALLIOVIRTA2016485}.

Financial returns are typically assumed to be uncorrelated or weakly correlated.
The stationary region of the parameters of MAR and MVAR models contains the uncorrelated
case, which allows these properties to be achieved smoothly as part of the estimation
process.
 
Using the Gaussian MVAR model assumption, we are able to fully specify
conditional predictive
distributions for future observations. We will show how it is possible to combine
modern portfolio theory \citep{Markowitz52} and the assumption of Gaussian 
mixture vector autoregressive model for portfolio optimisation. Under this 
model assumption, we will also estimate the risk associated with the forecast.
Finally, we will compare the performance of our method with that of the 
dynamic conditional correlation model by \citet{Engle2002} and the
vector autoregressive model (VAR).

\section{The mixture vector autoregressive model}
\label{sec:MVAR}
Mixture vector autoregressive models or MVAR \citep{Fong2007} are the 
multivariate extension of Mixture Autoregressive
Models \citep{WongLi2000}.

The MVAR model with $g$ Gaussian components, and an $m$ dimensional observation
vector $Y_t$
is defined as
\begin{equation}
F(Y_t \mid \mathcal{F}_{t-1}) = \sum_{k=1}^{g} \pi_k 
\Phi \left(\Omega_k^{-1/2} \left(Y_t - \Theta_{k0} - \sum_{i=1}^{p_k}
\Theta_{ki} Y_{t-i}\right)\right)
\label{eq:cdf}
\end{equation}
where
\begin{itemize}
\item $Y_t$ is a $m\times1$ data vector at time $t$.
\item \boldsymbol{$\pi$} = $\left(\pi_1, \ldots, \pi_g\right)$ are the mixing
  weights, such that $\pi_i > 0$ for $i=1,\dots,g$, and $\sum_{i=1}^{g}\pi_{i} = 1$.
\item $\Omega_k$ is the covariance matrix of component $k$.
\item $p_k$, $k=1,\ldots,g$ is the autoregressive order of component $k$. We 
denote $p=\max(p_k)$.
\item $\Theta_{k0}$ is a $m\times1$ intercept vector for component $k$, 
and $\Theta_{k1},\ldots,\Theta_{kp_k}$ are $m\times m$ matrices of autoregressive
parameters. If $p_k < p$, then $\Theta_{kl} = 0_m$ for $p_k < l \leq p$, where
$0_m$ is the zero-matrix of size $m \times m$.
\item $\Phi\left(\cdot\right)$ is the CDF of the standard multivariate Normal 
distribution, and $\phi\left(\cdot\right)$ is the corresponding pdf. 
\item Assuming start at $t=1$, \eqref{eq:cdf} holds for $t>p$.
\end{itemize} 
Regularity conditions and parameter estimation by EM algorithm are
discussed in \citet{Fong2007} and \citet{KALLIOVIRTA2016485}.

MVAR may be seen as an
alternative to multivariate GARCH when the data presents heteroskedasticity and
time-dependent correlation matrices, 
while also accounting for possible multimodality and
asymmetry in the distribution.

For parameter estimation, we recur to the missing data formulation.
Suppose that a m-variate time series $\lbrace Y_t \rbrace$ of length $n$ follows
a MVAR process. Let
$\boldsymbol{Z} = \left(\boldsymbol{Z}_1, \ldots, \boldsymbol{Z}_n\right)$ be
an unobserved allocation random variable, where $\boldsymbol{Z}_t$ is a 
g-dimensional vector with component $k$ equal to $1$ if $Y_t$ comes
from the $k^{th}$ component, and $0$ otherwise, and such that exactly one element
of $\boldsymbol{Z}_t$ is equal to $1$.

Following notation from \citet{Fong2007}, let $\tilde{\Theta}_k = \left[\Theta_{k0}, 
\Theta_{k1},\ldots, \Theta_{kp_k}\right]$ and $X_{tk} = \left(1, Y_{t-1}^T, 
\ldots, Y_{t-p_k}^T\right)^T$. In addition, let $\vartheta$ denote the 
complete set of parameters. Parameter estimates are then obtained by
EM-algorithm \citep{dempster1977maximum} with the following steps:
\begin{itemize}
\item \textbf{E-step}
\begin{equation}
\tau_{tk} = \Ex \left[Z_{tk} \mid Y_t, \vartheta \right] = \dfrac{
	\pi_k ~\boldsymbol{\phi} \left( \Omega_k^{-1/2} \left(Y_t - \Theta_{k0} - \displaystyle \sum_{i=1}^{p_k}
	\Theta_{ki} Y_{t-i}\right) \right)}{
	\displaystyle \sum_{l=1}^{g}
\pi_l ~\boldsymbol{\phi} \left( \Omega_l^{-1/2} \left(Y_t - \Theta_{l0} - \sum_{i=1}^{p_l}
\Theta_{li} Y_{t-i}\right) \right)}
\label{eq:Estep}
\end{equation}
\item \textbf{M- step}
\begin{equation}
\begin{split}
\hat{\pi}_k &= \dfrac{1}{n-p} \sum_{t=p+1}^{n} \tau_{tk} \\
\hat{\tilde{\Theta}}_k &= \left(\sum_{t=p+1}^{n} \tau_{tk} X_{tk}X_{tk}^T\right)^{-1}
\left(\sum_{t=p+1}^{n} \tau_{tk} X_{tk} Y_t^T\right) \\
\hat{\Omega}_k &= \dfrac{\displaystyle 
	\sum_{t=p+1}^{n}\tau_{tk}e_{tk}e_{tk}^T}{
	\displaystyle \sum_{t=p+1}^{n} \tau_{tk}}
\end{split}
\label{eq:Mstep}
\end{equation}
where $e_{tk} = Y_t - \Theta_{k0} - \sum_{i=1}^{p_k}
\Theta_{ki} Y_{t-i}$.
\end{itemize}
E-step and M-step are repeated recursively until convergence to maximum
likelihood estimates of the parameters.

First and second order stationarity conditions are discussed by
\citet{Saikkonen2007stability}, see also \citet{BOSHNAKOV2011415} for the
univariate case. Let
\begin{equation}
A_k = \begin{bmatrix}
\Theta_{k1} &\Theta_{k2} &\dots   &\Theta_{kp-1} &\Theta_{kp} \\
 I_m        & 0_m        &\dots   & 0_m          & 0_m \\ 
 0_m        & I_m        &\dots   & 0_m          & 0_m \\
\vdots      &\vdots      &\ddots  &\vdots        &\vdots \\
 0_m        & 0_m        &\dots   & I_m          & 0_m      
\end{bmatrix}, \qquad k=1,\ldots,g
\end{equation}
where $I_m$ and $0_m$ are respectively the identity matrix and
the zero matrix of size $m \times m$.
A necessary and sufficient condition for the MVAR model to be 
stationary is that the eigenvalues of 
$\sum_{k=1}^{g} \pi_k A_k \otimes A_k$ are smaller than $1$ in 
modulus. A MVAR model that satisfies this condition is said
to be \textit{Stable}. In practice, to assess stability of the
fitted model, parameters are replaced by their estimates.
\subsection{Prediction with mixture vector autoregressive models}
\label{sec:predMVAR}

In the context of mixture models, density forecasts are often more
attractive than point predictors and prediction intervals. This is because the
qualitative features of a predictive distribution, such as multiple modes or
skewness, are more intuitive and useful than simply a forecast 
and the associated prediction interval. In addition, when the predictive
distribution is available, prediction intervals can easily be obtained by
extracting the quantiles of interest
\citep{boshnakov2009mar, lawless2005pred}. 
Therefore, we here present derivation of
full predictive distributions for MVAR models, which will be used
throughout the analysis.

By model assumption, the one step ahead conditional predictive distribution
at time $t$ is fully specified, and it is that of \eqref{eq:cdf} where, for
notational convenience, we replace $t$ with $t+1$, i.e.  
\begin{equation*}
F(Y_{t+1} \mid \mathcal{F}_{t}) = \sum_{k=1}^{g} \pi_k 
\Phi \left(\Omega_k^{-1/2} \left(Y_{t+1} - \Theta_{k0} - \sum_{i=1}^{p_k}
\Theta_{ki} Y_{t+1-i}\right)\right)
.
\end{equation*}
%
Thus, the conditional distribution of the one step ahead predictor is a mixture
of $g$ Gaussian components and it depends on previous observations. In
particular, the conditional covariance matrix depends on previous values of the
process, a defining property of heteroskedasticity.
To obtain the conditional
mean and the covariance matrix let
$\mu_{t+1,k} = \Theta_{k0} +
\sum_{i=1}^{p_k} \Theta_{ki} Y_{t+1-i}$, for $k=1,\ldots,g$.
Then
\begin{equation} 
\begin{split}
\Ex \left[Y_{t+1} \mid \mathcal{F}_{t}\right] &= \sum_{k=1}^{g} \pi_k \mu_{t+1,k} = \mu_{t+1} \\
\Cov\left(Y_{t+1} \mid \mathcal{F}_{t}\right) &= \sum_{k=1}^{g}\pi_k \Omega_k + 
\sum_{k=1}^{g} \pi_k \left(\mu_{t+1,k} - \mu_{t+1} \right)
\left(\mu_{t+1,k} - \mu_{t+1} \right)^T \\
&= \sum_{k=1}^{g} \pi_k \Omega_k + \sum_{k=1}^{g} \pi_k \mu_{t+1,k}\mu_{t+1,k}^T
- \mu_{t+1} \mu_{t+1}^T
\end{split}
\label{eq:cond}
\end{equation}

Using a method analogous to that of \citet{boshnakov2009mar}, we can derive 
the conditional distribution for the two-step ahead predictor as a 
mixture of $g^2$ Gaussian components:
\begin{equation}
F\left(Y_{t+2} \mid \mathcal{F}_t\right) = \sum_{k=1}^{g} \sum_{l=1}^{g}
\pi_k \pi_l \Phi\left(\Psi_{kl}^{-1/2} \left(Y_{t+2} - \mu_{kl}\right)\right)
\label{eq:2stepdist}
\end{equation}
where, for each $k,l=1,\ldots,g$, 
\begin{equation*}
\begin{split}
\mu_{kl} &= \Theta_{k0} + \Theta_{k1}\Theta_{l0} +
\sum_{i=1}^{p-1} \left(\Theta_{k,i+1} + \Theta_{k1}\Theta_{li}\right) Y_{t - 1 - i}
+ \Theta_{k1} \Theta_{lp} Y_{t-1-p} \\
\Psi_{kl} &= \Omega_{k} + \Theta_{k1} \Omega_{l} \Theta_{k1}^T
\end{split}
\end{equation*}
Note that, in general, $\mu_{kl} \neq \mu_{lk}$ and $\Psi_{kl} \neq \Psi_{lk}$. 
Expectation and covariance matrix of this predictor are:
\begin{equation} 
\begin{split}
\Ex \left[Y_{t+2} \mid \mathcal{F}_{t}\right] &= \sum_{k=1}^{g}\sum_{l=1}^{g} \pi_k \pi_l \mu_{kl} = \mu_{t+2} \\
\Cov\left(Y_{t+2} \mid \mathcal{F}_{t}\right) &= \sum_{k=1}^{g} \pi_k \pi_l \Psi_{kl} 
+ \sum_{k=1}^{g} \sum_{l=1}^{g} \pi_k\pi_l \mu_{kl}\mu_{kl}^T
- \mu_{t+2} \mu_{t+2}^T
\end{split}
\label{eq:cond2}
\end{equation}
Full derivation of 
\eqref{eq:cond2}, as well as proof of the
conditional distribution of $Y_{t+2}$, is available in Appendix~\ref{app:derive}. By
recursing this procedure, we could derive a full distribution for any horizon
$h$. However, the number of components in the mixture increases to $g^{h}$ when
$h$ increases and therefore simulation methods may be preferred for approximate
computation of predictive densities for larger horizons.


\section{Portfolio optimisation with MVAR models}
\label{sec:mvarportfolio}

Suppose that a multivariate time series $\lbrace Y_t \rbrace$ of asset returns 
is observed, and it is
believed that the underlying generating process is MVAR. 
From Section \ref{sec:predMVAR},
conditional distributions of the 1 and 2 step predictors are fully specified, and
can be estimated by plugging parameter estimates into the relevant equations.

Now, let $w$ denote the weights of a portfolio built with assets $\lbrace Y_t \rbrace$
 (allowing
short selling), and let
$R_{t+1} = w^TY_{t+1}$ be the portfolio return at time $t+1$.  
Intuitively, because our model consists of a mixture of multivariate 
normal components, we can apply the property in \eqref{eq:multtouniv} to
conclude that the conditional distribution of $R_{t+1}$ is also (univariate) mixture normal,
with corresponding mixing weights $\boldsymbol{\pi}$ from the fitted
multivariate model. By model assumption in fact, at each time $t+1$ an observation
 $Y_{t+1}$ is assumed to be generated from one of $g$ components of the mixture.
 Consequently, $R_{t+1}$ is obtained by applying \eqref{eq:multtouniv} to the
 selected component. Recursing this for all $g$ components  
 the result is itself a mixture distribution for $R_{t+1}$. 
 
 In terms of MVAR model parameters we write:
\begin{equation}
F(R_{t+1} \mid \mathcal{F}_{t}) = \sum_{k=1}^{g} \pi_k \Phi\left(
\dfrac{R_{t+1} - w^T \mu_{t+1,k}}{\sqrt{w^T \Omega_k w}}\right)
\end{equation}
Conditional mean and variance of $R_{t+1}$ are:
\begin{equation}
\begin{split}
\Ex[R_{t+1} \mid \mathcal{F}_{t}] &= \sum_{k=1}^{g} \pi_k \left(w^T \mu_{t+1,k}\right) 
= \sum_{k=1}^{g} \pi_k \mu_{t+1,k}^{*} = \mu^{*} \\
\Var(R_{t+1} \mid \mathcal{F}_{t}) &= \sum_{k=1}^{g}\pi_k \left(w^T \Omega_k w \right)
+\sum_{k=1}^{g}\pi_k \left(\mu_{t+1,k}^{*}\right)^2 - \left(\mu^{*} \right)^2  
\end{split}
\label{eq:sumstats}
\end{equation}

Modern portfolio theory \citep{Markowitz52} gives us a way to calculate 
weights $w^{*}$ to construct the 
most
efficient portfolio for a given return, and to calculate the efficient portfolio 
of assets with the minimum possible variance. A portfolio with target return $\mu$
is said to be an efficient portfolio when the variance associated with it is the
lowest amongst all portfolios of the same assets having that same
target return. The minimum variance portfolio is the efficient portfolio with the
lowest possible variance of all efficient portfolios of the same assets.
For the remainder of the analysis, we will denote efficient portfolios with 
the subscript $\EFF$,
and minimum variance portfolios with the subscript $\MVP$.
We now see how modern portfolio theory can be used to predict future observations 
assuming a MVAR model.

For the MVAR case, let $\Ex\left[Y_{t+1} \mid \mathcal{F}_{t}\right]=\mu_{t+1}$ and
$\Cov\left(Y_{t+1} \mid \mathcal{F}_{t}\right) = \Omega_{t+1}$.
In addition, let
\begin{equation}
A = \mathbbm{1} \Omega_{t+1}^{-1} \mu_{t+1} \quad, \quad
B = \mu_{t+1} \Omega_{t+1}^{-1} \mu_{t+1} \quad , \quad
C = \mathbbm{1} \Omega_{t+1}^{-1} \mathbbm{1} \quad , \quad
D = CB - A^2
\label{eq:ABCD}
\end{equation}
where $\mathbbm{1}$ is a vector of $1$s of the same length as $\mu_{t+1}$.

It can be proved that optimal weights for an efficient portfolio of these assets 
and target return $\mu_{\EFF}$ are
\begin{equation}
w_{\EFF} = \dfrac{1}{D} \left(B \Omega_{t+1}^{-1} \mathbbm{1} - 
A \Omega_{t+1}^{-1} \mu_{t+1} + \mu^{*}\left(
C \Omega_{t+1}^{-1}\mu_{t+1} - A \Omega_{t+1}^{-1} \mathbbm{1}\right)\right)
\label{eq:wstar}
\end{equation}
and the variance of such portfolio can be calculated equivalently as 
$Var(R_t \mid \mathcal{F}_{t-1})$ (MVAR model assumption) or 
$w^T \Omega_t w$ (the variance of an efficient
portfolio of assets) since
\begin{equation}
\begin{split}
w_{ \EFF}^{T} \Omega_{t+1} w_{\EFF} &= 
\sum_{k=1}^{g}\pi_k \left(w_{\EFF}^{T} \Omega_k w_{\EFF} \right)
+\sum_{k=1}^{g}\pi_k \left(w_{\EFF}^{T} \mu_{t+1,k}\right)^2 \\ &- 
\left[\sum_{k=1}^{g} \pi_k \left(w_{\EFF}^{T} \mu_{t+1,k} \right)\right]^2 = Var(R_{t+1} \mid \mathcal{F}_{t})
\end{split}
\label{eq:EFFvar}
\end{equation}
In practice, $\mu_{t+1,k}$ and $\Omega_{t+1}$ are replaced with their estimates
$\hat{\mu}_{t+1,k}$ and $\hat{\Omega}_{t+1}$.

Weights of the minimum variance
portfolio of same assets $\lbrace Y_t \rbrace$, and corresponding return, are:
\begin{equation}
w_{\MVP} = \dfrac{\Omega_{t+1}^{-1} \mathbbm{1}}{C} \qquad \mu_{\MVP} = \dfrac{A}{C}
\label{eq:mvp}
\end{equation}
Conditional predictive distributions can also be calculated analytically for
any $h\geq2$. However, one must keep in mind that such predictive
distribution would be a mixture of $g^h$ components, so that simulation 
methods may be preferred in some cases as $h$ increases.

Consider the case $h=2$. The conditional predictive distribution 
$F\left(Y_{t+2} \mid \mathcal{F}_t\right)$ for the MVAR model 
is a mixture of $g^2$ Gaussian
components given in \eqref{eq:cond2}.
Similarly to the case $h=1$, we can derive the full conditional distribution 
of $R_{t+2}$, which is again a mixture of $g^2$ Gaussian components:
\begin{equation}
F(R_{t+2} \mid \mathcal{F}_t) = \sum_{k,l=1}^{g} \pi_k \pi_l \Phi 
\left(\dfrac{R_{t+2} - w^{(2)}\mu_{kl}}{w^{(2)^T}\Psi_{kl} w^{(2)}} \right)
\label{eq:2stepreturn}
\end{equation}
where $w^{(2)}$ is the vector of optimal weights for this portfolio.
Similarly to the case $h=1$, one can now calculate 
$\Ex\left[Y_{t+2} \mid \mathcal{F}_{t}\right]=\mu_{t+2}$ and
$\Cov\left(Y_{t+2} \mid \mathcal{F}_{t}\right) = \Omega_{t+2}$ and adapt 
\eqref{eq:ABCD}, \eqref{eq:wstar}, \eqref{eq:EFFvar} and \eqref{eq:mvp} to 
obtain an efficient or minimum variance portfolio.

\section{Simulation example}
We simulate a series of size $n=500$ of hypothetical 
stock returns from 
the $3-$variate
 $\MVAR(2;1,1)$ process
\begin{equation*}
	F(Y_t \mid \mathcal{F}_{t-1}) = 
	0.75 \boldsymbol{\Phi}\left(\dfrac{Y_t - v_1}{\Omega_1}\right) 
	+ 0.25 \boldsymbol{\Phi}\left(\dfrac{Y_t - v_2}{\Omega_2}\right)
\end{equation*} 
where
\begin{equation*}
v_1 = \Theta_{10} + \Theta_{11}Y_{t-1} \qquad v_2 =  \Theta_{20} + \Theta_{21}Y_{t-1}
\end{equation*}
and
\begin{align*}
\Theta_{10} &= \begin{bmatrix}
0 \\ 0 \\ 0 \\ 
\end{bmatrix}
&\Theta_{11} = \begin{bmatrix}
0.5 &0 &0.4 \\
-0.3 &0 &0.5 \\
-0.6 &0.5 &-0.3
\end{bmatrix}
&\Omega_1 = \begin{bmatrix}
1 &0.5 &-0.40 \\
0.5 &2 &0.8 \\
-0.4 &0.8 &4 \\ 
\end{bmatrix}
\\
\Theta_{20} &= \begin{bmatrix}
0 \\ 0 \\ 0 \\ 
\end{bmatrix}
&\Theta_{21} = \begin{bmatrix}
-0.5 &1 &-0.4 \\
0.3 &0 &-0.2 \\
0 &-0.5 &0.5 \\ 
\end{bmatrix}
&\Omega_2 = \begin{bmatrix}
1 &0.2 &0 \\
0.2 &2 &-0.55 \\
0 &-0.55 &4
\end{bmatrix}
\end{align*}

\begin{figure}
\centering
\includegraphics[scale=0.4]{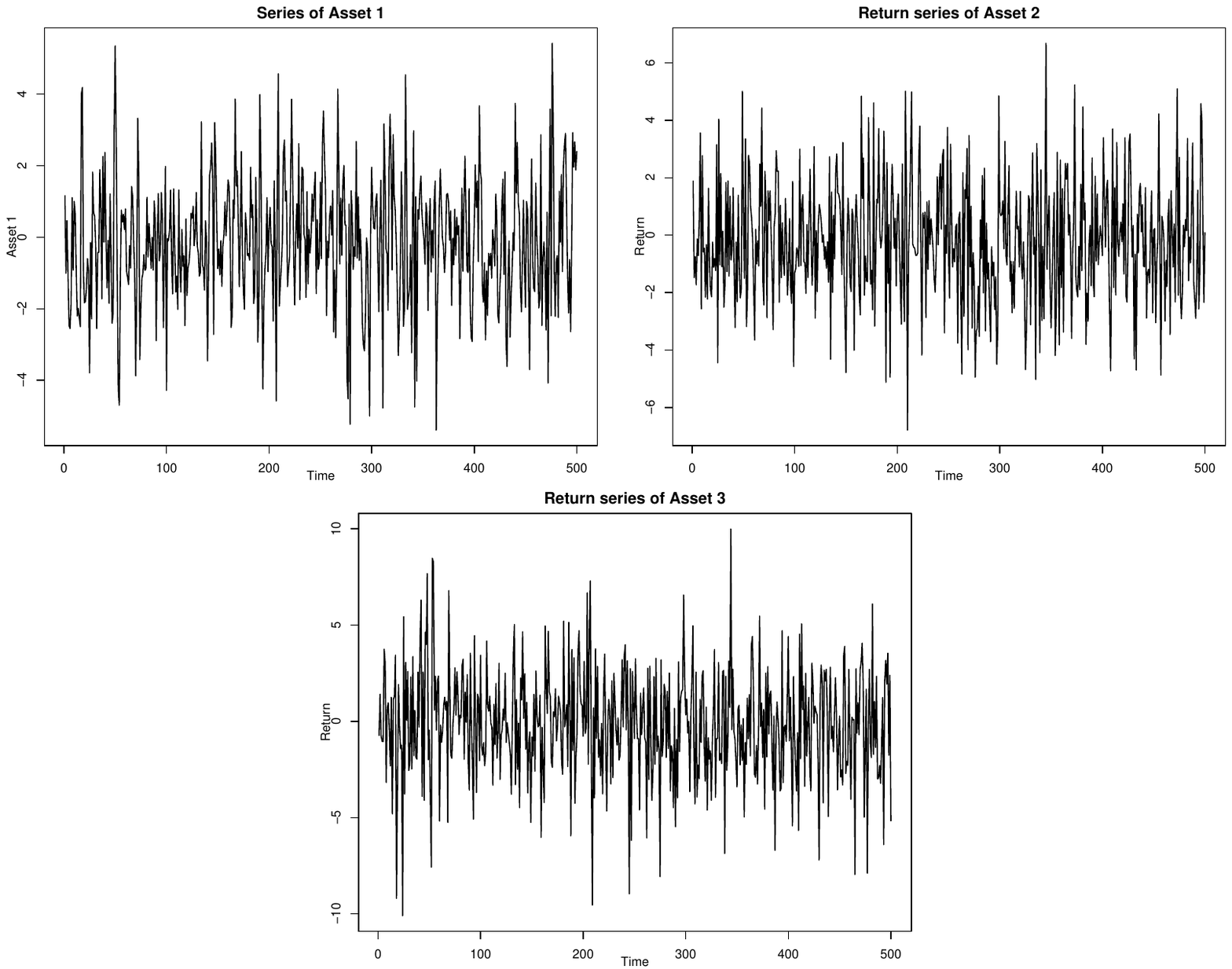}
\caption{Simulated time series of stock returns Asset 1 (top left), Asset 2 (top
	right) and Asset 3 (bottom).}
\label{fig:simTS}
\end{figure}

\begin{figure}
\centering
\includegraphics[scale=0.4]{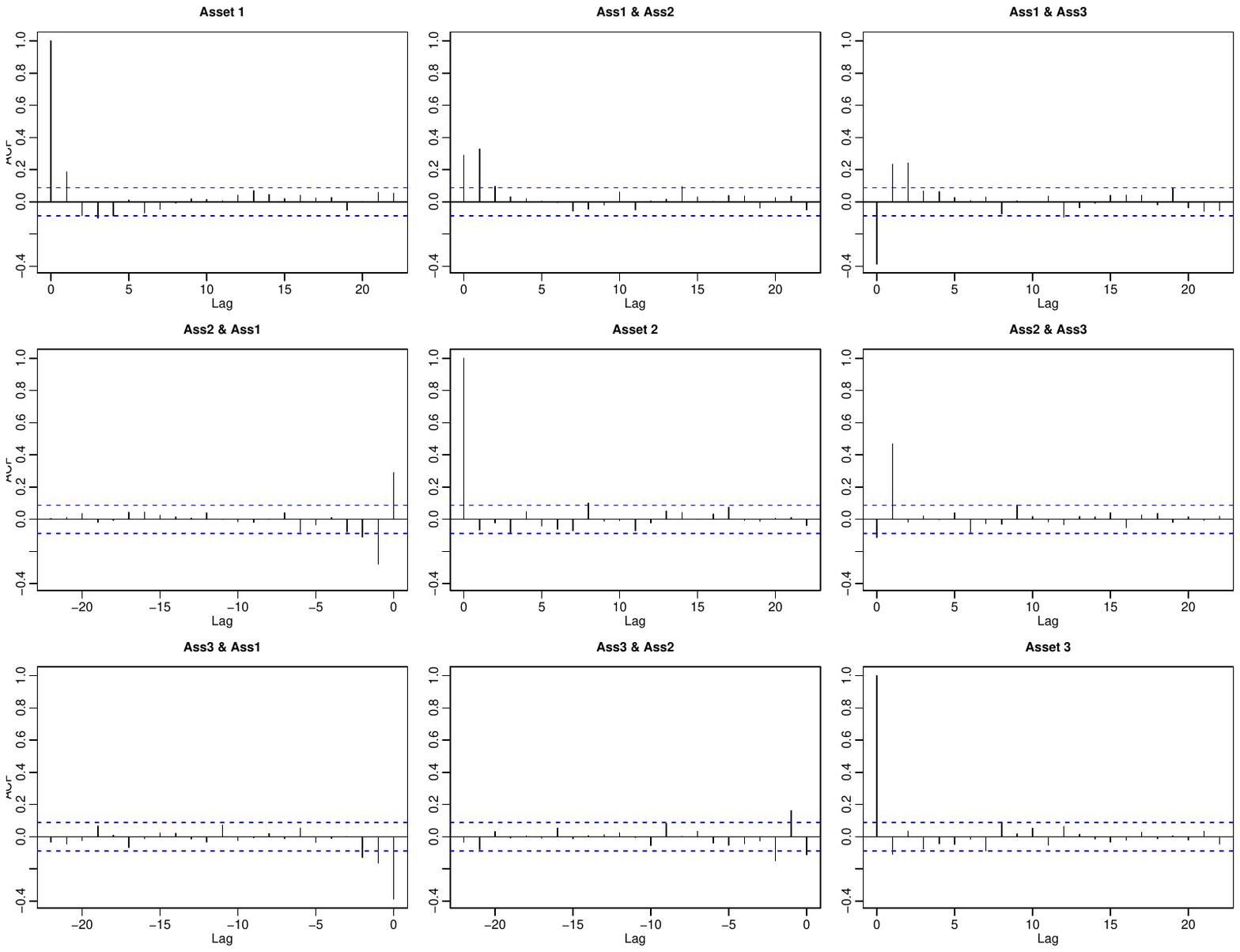}
\caption{Autocorrelation and corss-correlation plots of the simulated time
series data.}
\label{fig:simACF}
\end{figure}

The three univariate series can be seen in Figure \ref{fig:simTS}, 
with their autocorrelation and
cross-correlation plots in Figure \ref{fig:simACF}. The data is very representative 
of what we should be looking for, in a real case scenario, to assume an underlying
MVAR process. We notice in fact signs of heteroskedasticity in each of the
 series, 
and autocorrelations and cross-correlations significantly different from $0$
at lags larger than $0$. The latter is what separates MVAR from multivariate
GARCH models, which assume the original series to be uncorrelated.


Parameter estimates were calculated using the EM Algorithm with the formulas in
\eqref{eq:Estep} and \eqref{eq:Mstep}. 
In order to perform out of sample prediction, data from $Y_1$
to $Y_{498}$ were used for estimation, with $Y_{499}$ and $Y_{500}$ being left
out as observations 1 and 2 time points in the future:

\begin{gather*}
  \begin{aligned}
    \hat{\boldsymbol{\pi}}
    &= (0.7242, 0.2758)
  \end{aligned}
  \\
  \begin{aligned}
    \Theta_{10}
    &= \begin{bmatrix}
      -0.0022 \\ -0.0303 \\ 0.1276 \\ 
    \end{bmatrix}
    &\hat{\Theta}_{11} =
    \begin{bmatrix}
      0.4931 &-0.0339 &0.4169 \\
      -0.3156 &-0.0012 &0.5078 \\
      -0.6141 &0.6007 &-0.3844
    \end{bmatrix}
    &\hat{\Omega}_1 =
    \begin{bmatrix}
      0.9551 &0.4783 &-0.2776 \\
      0.4783 &1.9123 &0.9736 \\
      -0.2776 &0.9736 &3.9455
    \end{bmatrix}
    \\
    \hat{\Theta}_{20}
    &= \begin{bmatrix}
      0.0338 \\ 0.5499 \\ -0.7580
    \end{bmatrix}
    &\hat{\Theta}_{21} =
    \begin{bmatrix}
      -0.4595 &1.0124 &-0.4004 \\
      0.3343 &-0.1423 &-0.1551 \\
      -0.1273 &-0.2336 &0.6509
    \end{bmatrix}
    &\hat{\Omega}_2 =
    \begin{bmatrix}
      0.8767 &0.4794 &-0.3627 \\
      0.4794 &2.9148 &-0.6576 \\
      -0.3627 &-0.6576 &9.8135
    \end{bmatrix}
  \end{aligned}
\end{gather*}

We then calculated the one step ahead conditional mean and variance based on
parameter estimates:
\begin{equation*}
  \begin{split}
    \Ex\left[Y_{499} \mid \mathcal{F}_{498}\right] &= \hat{\mu}_{499} =
    \begin{bmatrix}
      -0.1750 \\
      -0.9655  \\
      -1.4361
    \end{bmatrix}
    \\
    \Cov(Y_{499} \mid \mathcal{F}_{498}) &= \hat{\Omega}_{499}  =
    \begin{bmatrix}
      1.3109 &-0.6080 &-0.0768 \\
      -0.6080 &5.3174 &-0.5642 \\
      -0.0768 &-0.5642 &5.9420 
    \end{bmatrix}
  \end{split}
\end{equation*}

Given $\hat{\Omega}_{499}$, we can calculate the minimum variance portfolio, which is
obtained for weights $w_{\MVP} = (0.6434, 0.2228, 0.1338)^T$. The corresponding
expected return on this portfolio at $(t+1)=499$ is $\mu_{\MVP} = -0.5198$, with
standard deviation $\sigma_{\MVP} = 0.8475$.

Suppose now that we wish to increase our return to $\mu^{*} = 0$, i.e. no expected
 loss, at the cost of a larger variance. We can calculate weights to construct an 
efficient portfolio of these
assets as seen in Section \ref{sec:mvarportfolio}.
We obtain:
\begin{equation*}
w_{\EFF} = \begin{bmatrix}
1.1097\\
0.0781\\
-0.1878
\end{bmatrix}
\end{equation*}
The interpretation of $w_{\EFF}$ is that the otpimal portfolio yielding expected 
return of $0$ is constructed by short-selling a small amount of Asset 3, and
investing $110.97\%$ and $7.81\%$ of the initial capital (meanwhile increased by
short selling) into Asset 1 and Asset 2
respectively.
Notice that the target return is $w_{\EFF} \mu_t = \mu_{\EFF} = 0$ as desired.

We can now calculate the quantities we need for the conditional predictive 
distribution of $R_{499}$:
\begin{align*}
&\mu_1^{*} = w^{*}\mu_{499,1} = 0.2642 
&\sigma_1^{*^{2}} = 1.4968 \approx (1.2235)^2 \\
&\mu_2^{*} = w^{*}\mu_{499,2} = -0.6939 
&\sigma_2^{*^{2}} = 1.6062 \approx (1.3025)^2
\end{align*}
Therefore, the conditional distribution of $R_{499} = w^{*^{T}} Y_{499}$ is
\begin{equation*}
F(R_{499}\mid \mathcal{F}_{498}) = 0.7242 \times \Phi\left(\dfrac{R_{499} - 0.2642}{1.2235}\right) +
0.2758 \times \Phi\left(\dfrac{R_{499} - 0.6939}{1.3025}\right)
\end{equation*}
 
\begin{figure}
\centering
\includegraphics[scale=0.4]{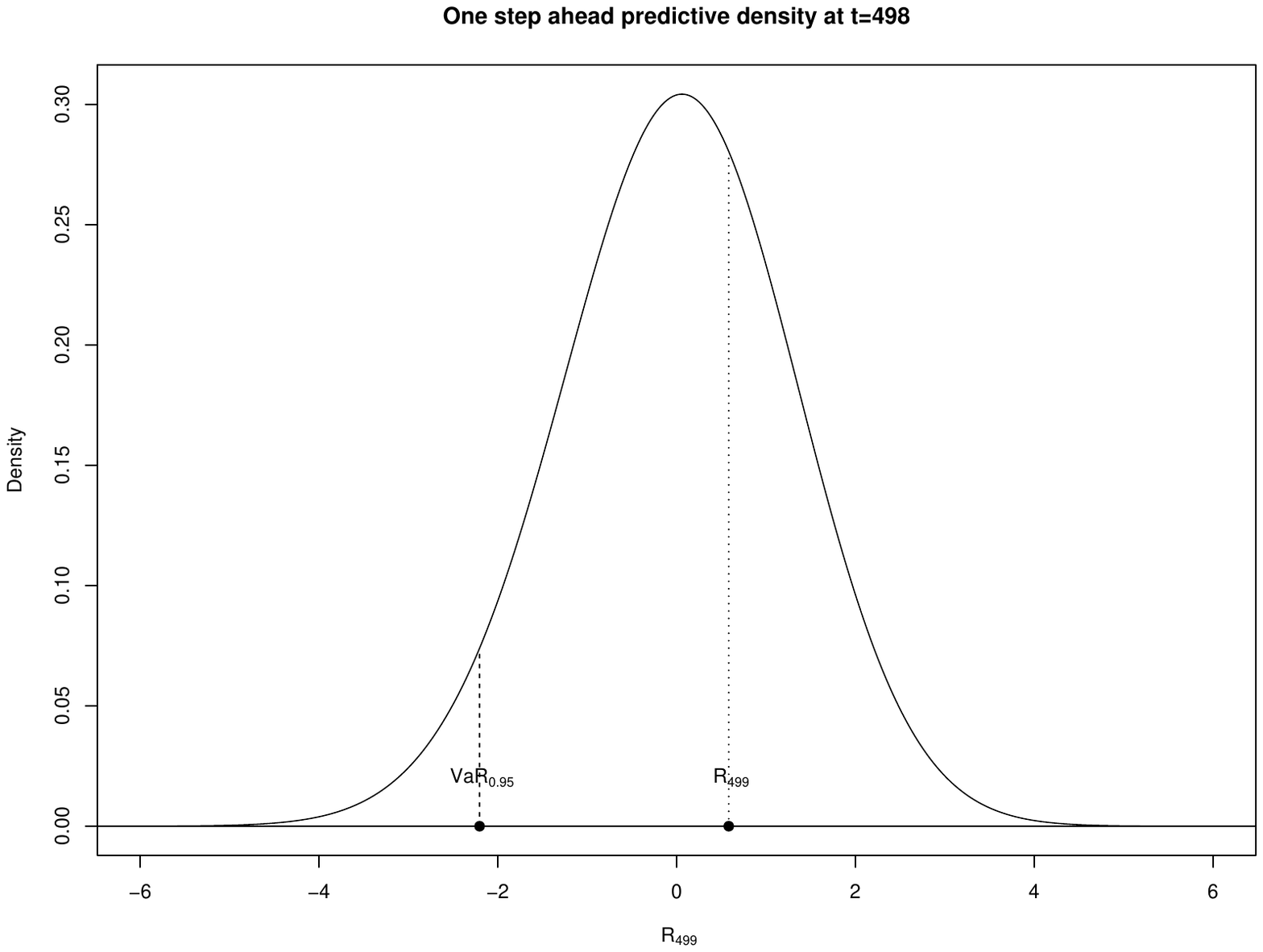}
\caption{Conditional one-step predictive density
of $R_{499}$, with VaR at $95\%$ (dashed line) and observed return 
(dotted line) highlighted.}
\label{fig:rEFF}
\end{figure}

The standard deviation associated to this portfolio
 is $\sigma_{\EFF} = 1.3173$ which as expected is larger than $\sigma_{\MVP}$. 
 More importantly,
we can use the distribution assumption on $R_{499}$ to estimate risk measures.
 Figure \ref{fig:rEFF} shows the conditional
distribution of $R_{499}$. The dot on the left hand side, highlighted with
a dashed line, is
the value at risk at $95\%$ level. It was found that the value
at risk at such level is $-2.2039$, with expected shortfall of $-2.7912$.
This means that an investor could expect a loss on this portfolio higher than
 $2.2039$ with probability $0.05$, and when this threshold is exceeded, 
 the expected loss is of $2.7912$.
 The observed return is also shown in Figure \ref{fig:rEFF} 
 as a dot with dotted line. We notice that it lies on a region of
 high density of the predictive distribution.

\begin{figure}
\centering
\includegraphics[scale=0.4]{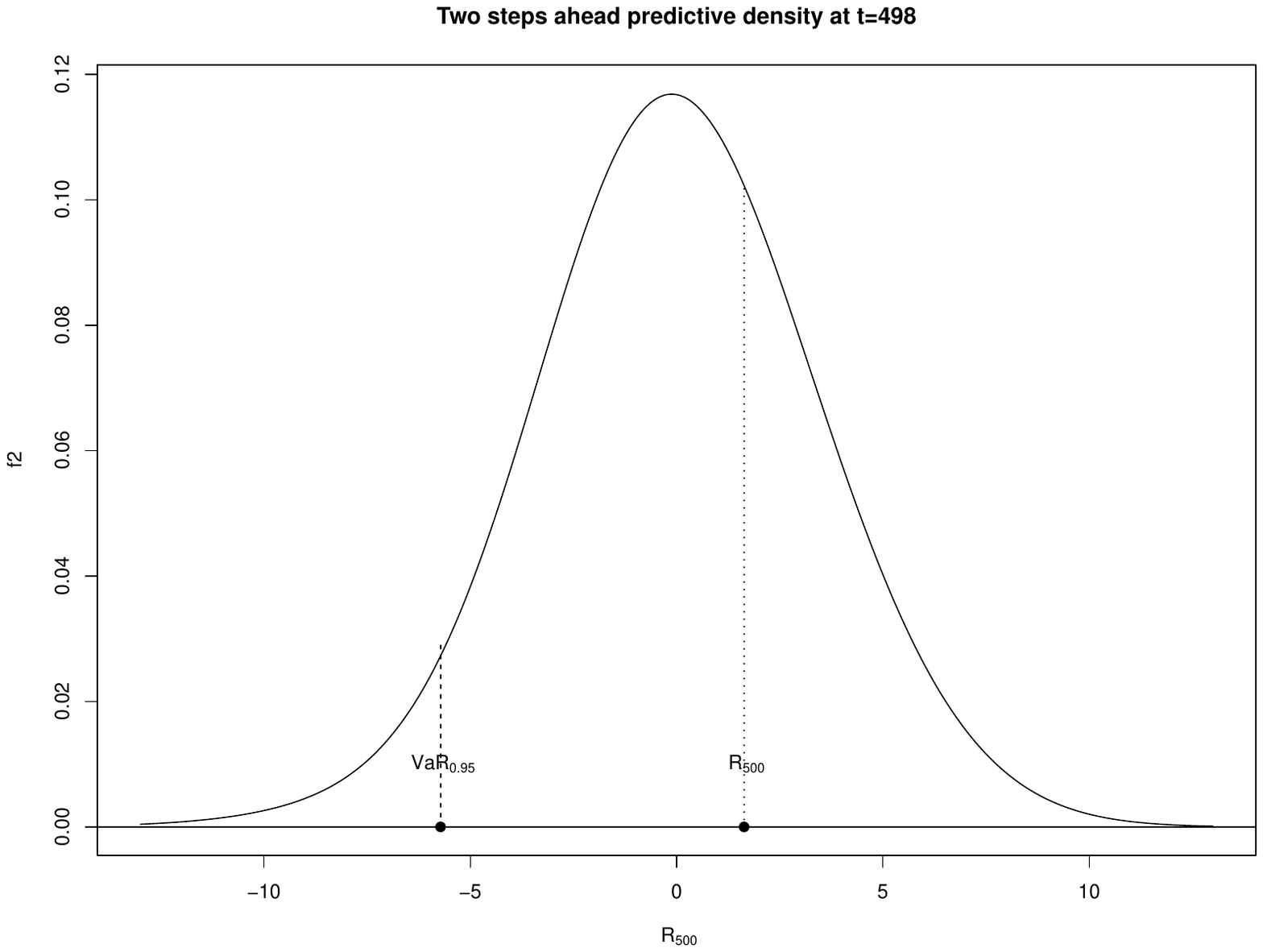}
\caption{Conditional two-step predictive density
	of $R_{500}$, with VaR at $95\%$ (dashed line) and observed return 
	(dotted line) highlighted.}
\label{fig:rEFF_pred2}
\end{figure}

We can also estimate the conditional distribution of the two-step ahead 
predictor at $t=498$,
 $F(R_{500} \mid \mathcal{F}_{498})$. This isshown in Figure
  \ref{fig:rEFF_pred2}).

The minimum variance portfolio for a a two-step ahead portfolio of assets is
calculated with weights $w_{\MVP}^{(2)} = (0.4367, 0.2822, 0.2811)$, with an
expected return $\mu_{\MVP}^{(2)}=-0.3918$, with $\sigma_{\MVP}^{(2)} = 1.1784$,
showing the increasing uncertainty as we attempt to predict further into the
future.

Once again we consider building a portfolio of assets yielding expected 
return $\mu^{*}=0$. Optimal weights for this portfolio are 
\begin{equation*}
w_{\MVP}^{(2)} =\begin{bmatrix}
-0.9404 \\
1.5193 \\
0.4211 \\
\end{bmatrix}
\end{equation*}

From Figure \ref{fig:rEFF_pred2}, 
we notice how the density is now flatter, which is sign of a
larger variability. In fact, the estimated standard deviation of $R_{500}$ is
$3.5056$, which is a significant increase. 
This also results in much larger estimated
$\VaR = -5.0207$ (in absolute value) at the same $95\%$ level, 
with expected shortfall equal to $-7.4505$.
Once again, the observed return (dotted line) is in a high density 
region of the predictive distribution.

Overall, we can be satisfied with the performance of our method in predicting
portfolio returns.

\section{Application to the US stock market}
\label{sec:realdata}

\begin{figure}[h]
\centering
\includegraphics[scale=0.5]{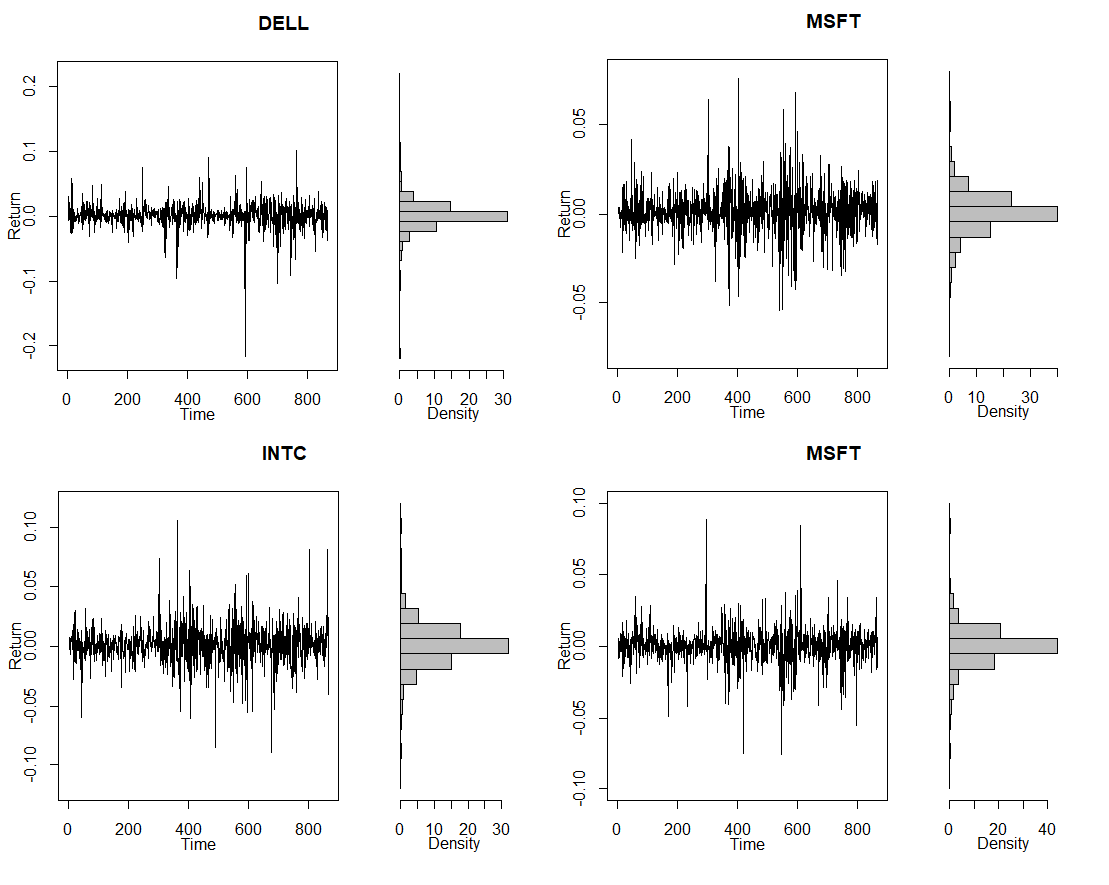}
\caption{Time series of returns of DELL (top left), 
	\\	MSFT (top right), INTC (bottom left) and IBM (bottom right).}
\label{fig:realdata}
\end{figure}

We consider a multivariate dataset of $m=4$ stocks on the US stock market:
Dell Technologies Inc. (\texttt{DELL}),
Microsoft Corporation (\texttt{MSFT}),
Intel Corporation (\texttt{INTC}),
and
International Business Machine Corporation (\texttt{IBM}).
The data were obtained from Yahoo! Finance (\url{https://finance.yahoo.com}). 
The original time series include daily Adjusted Close Prices between January 
2$^{nd}$ 2016 and January 29$^{th}$ 2020 (867 observations).
For each series and $t=2,\ldots,867$, we calculated daily returns as 
$(\text{Price}_t - \text{Price}_{t-1})/\text{Price}_{t-1}$.
The resulting series, displayed in Figure \ref{fig:realdata}, 
includes $866$ observations.

\begin{figure}
	\centering
	\includegraphics[scale=0.5]{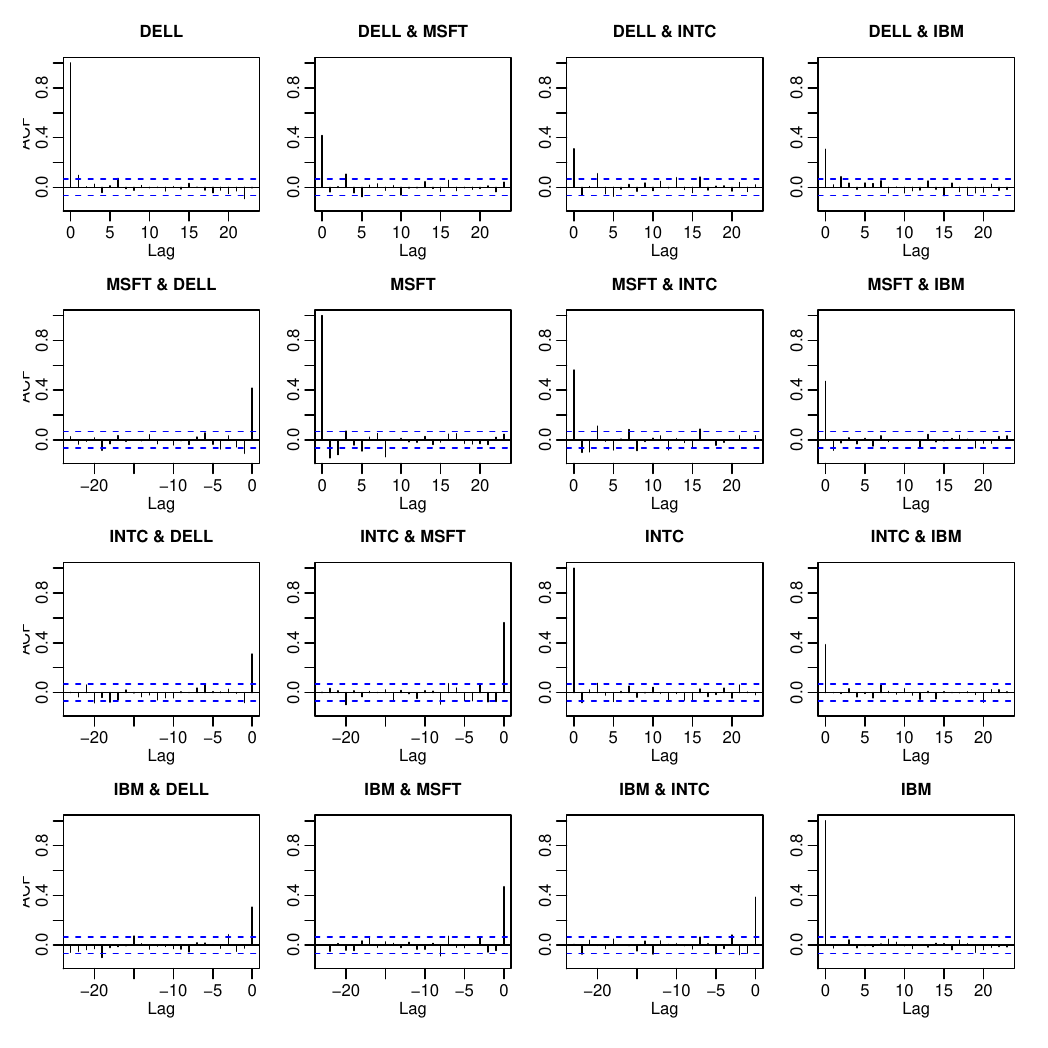}
	\caption{Autocorrelation and cross-correlation plots for the multivariate
		time series. Notice the presence of correlation and cross-correlation in
		the data.}
	\label{fig:acfrealdata}
\end{figure}

All four univariate
series in Figure \ref{fig:realdata} appear to be heteroskedastic. Their
histograms also show signs of heavy tails, which was confirmed by calculation of
sample excess kurtosis (all significantly larger than $0$).  In addition, from a
preliminary analysis, it was noticed that the data presents autocorrelation at
least at lags 1 and 2, and cross-correlations at lags 0, 1 and 2 (see Figure
\ref{fig:acfrealdata}). Therefore, it is reasonable to consider a MVAR generating
process for the data.

Several models were fitted. In terms of diagnostics, a $\MVAR(3;3,2,1)$ was chosen 
as best fit. 
Estimation was carried out on the first $864$ observations, omitting the last two
for out-of-sample prediction.

Given parameter estimates and the one-step ahead predictive distribution at $t=864$,
we calculate weights for the minimum variance portfolio built with these
assets, which yields a mean return of approximately $0.0024\, (0.24\%)$. The standard
deviation associated with this portfolio is $\sigma_{\MVP} = 0.0092$. 

Now, assume we would like to increase our mean return to $0.007 = 0.7\%$. We 
can calculate optimal weights 
\begin{equation*}
w_{\EFF} = \left(-0.5832,
0.9538,
0.1085,
0.5209\right)^T.
\end{equation*}
Weights are interpreted as follows: an investor shall short-sell an amount
of around
 $0.58$ times their inital capital in DELL stocks, and reinvest the new total in the
remaining three assets, with a major bet on MFST and IBM. The idea behind this is 
that it is believed that DELL stocks will decrease in value between the present
and the nearest future, and therefore one could short-sell to make a profit.
On the other hand, it is believed that the remaining three assets will increase
their value in the same time span, and in particular MSFT stocks.
However, the standard deviation associated with this portfolio is 
$\sigma_{\EFF} = 0.0139$, a slight increase compared to $\sigma_{\MVP}$, 
considering the scale of the data.

\begin{figure}
	\centering
	\includegraphics[scale=0.4]{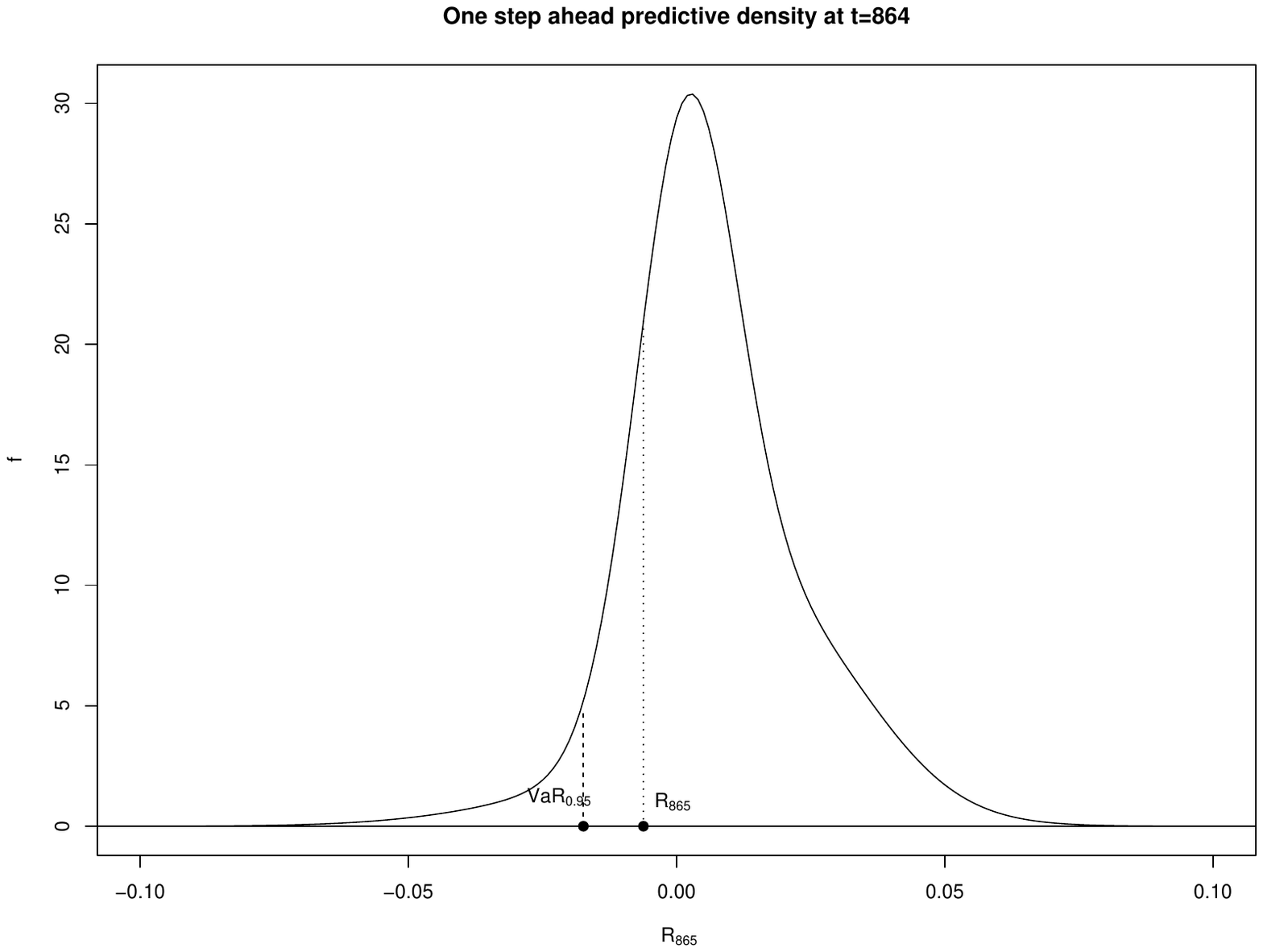}
	\caption{Conditional one-step predictive density
		of $R_{865}$, 
		with VaR at $95\%$ (dashed line) and observed return (dotted line)
		 highlighted.}
		\label{fig:dens1step}
	\end{figure}

For the latter portfolio, we calculate the one-step ahead conditional 
distribution
of $R_{865} = \sum_{m=1}^{4} w_m^{*} y_{m,865}$ using parameter estimates from 
the MVAR model fitting:
\begin{equation*}
\begin{split}
F(R_{865} \mid \mathcal{F}_{864}) = 
&0.1316 \Phi\left(\dfrac{R_{865}  + 0.00052 }{0.0266}\right) +
0.5627 \Phi\left(\dfrac{R_{865}  + 0.00178 }{0.0093}\right) \\+
&0.3057 \Phi\left(\dfrac{R_{865}  + 0.01932 }{0.0169}\right)
\end{split}
\end{equation*}
The corresponding predictive density can be seen in Figure \ref{fig:dens1step}.

Value at risk at $\alpha=95\%$ is estimated at $-0.0174$, with
expected shortfall of $-0.0299$. The subsequently observed return  is 
$R_{865} = -0.0062$, 
which we can see lies on a region of high density, 
and therefore is somewhat plausible.

We can also look at building a portfolio of the same assets looking two steps
into the future, at $t=866$. The minimum variance portfolio in this case yields
an expected return $\mu_{\MVP}^{(2)}=-0.011$, with associated standard deviation
$\sigma_{\MVP}^{(2)}=0.0101$. 
We use the distribution assumptions for $Y_{866}$,
its expected value and covariance matrix to estimate optimal weights to look once again to
increasing our return by bulding
an efficient portfolio with same target 
return $\mu^{*} = 0.007$ as before:
\begin{equation*}
w_{\EFF}^{(2)} = \left(
0.0402,
0.6174,
0.7554,
-0.4130
\right)^T
\end{equation*}

Using parameter estimates and \eqref{eq:2stepreturn}, the conditional
distribution of $R_{866}$ is a mixture of $3^2=9$ components:
\begin{equation*}
\begin{split}
F(R_{866} \mid \mathcal{F}_{864}) = 
&0.0173 ~\Phi\left(\dfrac{R_{866} - 0.0170}{0.0285}\right) 
+ 0.0741 ~\Phi\left(\dfrac{R_{866} -  0.0190}{0.0268}\right) \\
+ &0.0402 ~\Phi\left(\dfrac{R_{866} - 0.0252}{0.0276}\right) 
+ 0.0741 ~\Phi\left(\dfrac{R_{866} - 0.0086}{0.0101}\right) \\
+ &0.3166 ~\Phi\left(\dfrac{R_{866} - 0.0069}{0.0096}\right) 
+ 0.1720 ~\Phi\left(\dfrac{R_{866} - 0.0051}{0.0096}\right) \\
+ &0.0402 ~\Phi\left(\dfrac{R_{866} - 0.0098}{0.0206}\right) 
+ 0.1720 ~\Phi\left(\dfrac{R_{866} - 0.0040}{0.0187}\right) \\
+ &0.0935 ~\Phi\left(\dfrac{R_{866} + 0.0077}{0.0196}\right) 
\end{split}
\end{equation*}

\begin{figure}
	\centering
	\includegraphics[scale=0.4]{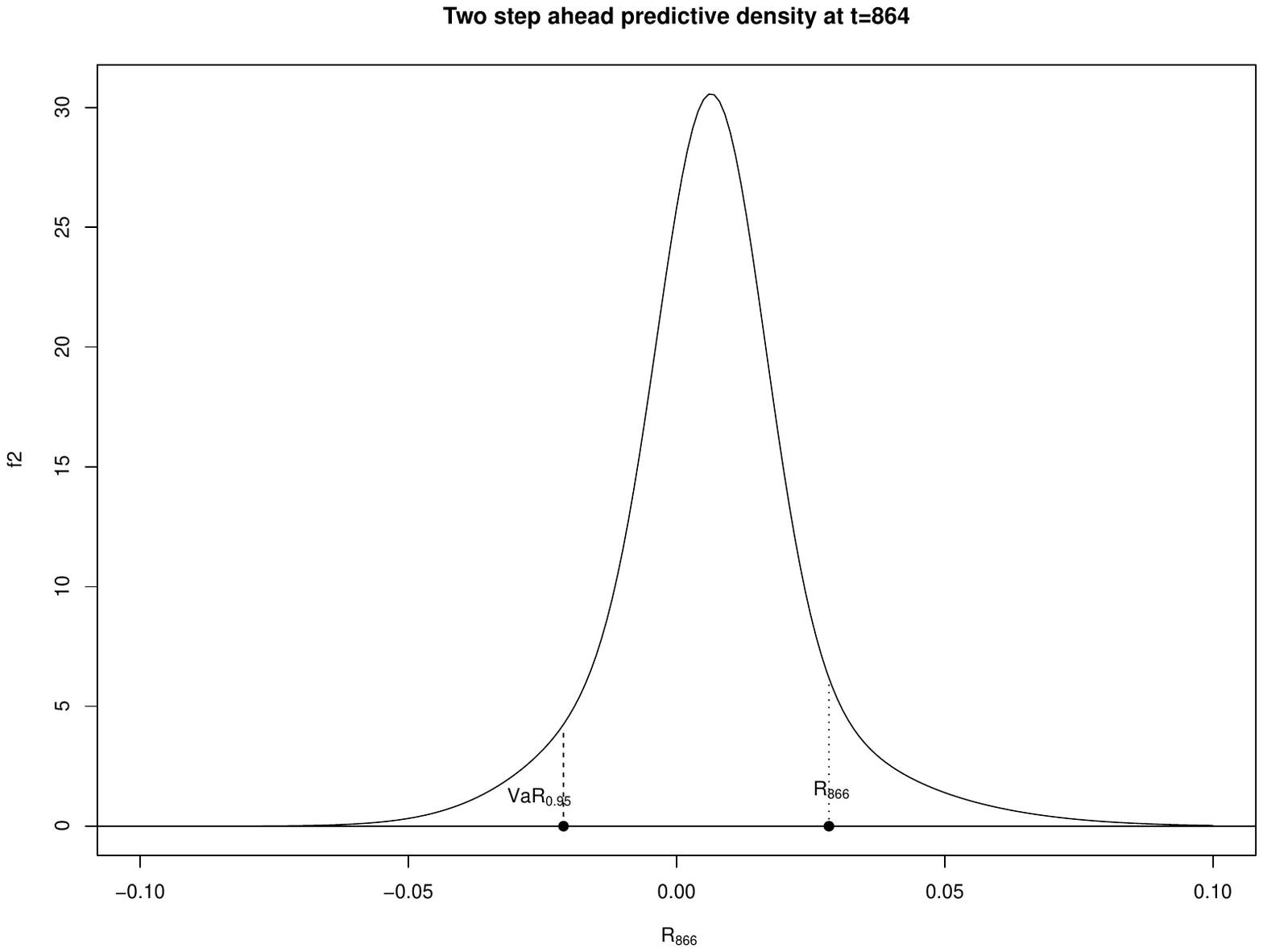}
	\caption{Conditional two-step predictive density
		of $R_{866}$, 
		with VaR at $95\%$ (dashed line) and observed return (dotted line)
		 highlighted.}
	\label{fig:dens2step}
\end{figure}

The conditional distribution of $R_{866}$ can be seen in Figure \ref{fig:dens2step}.
We notice, as expected, an increase in the standard deviation
of the distribution, 
$\sigma_{\EFF}^{(2)} = 0.0177$ with respect to $\sigma_{\EFF} = 0.0177$. 
Overall, the two shapes in Figure \ref{fig:dens1step} and \ref{fig:dens2step} look
 similar, however the observed return $R_{866}$ is not in a high density region of
 its predictive distribution, as it actually exceeds the expectations.
$\VaR$ is now estimated at $-0.021$, 
with expected shortfall equal to $-0.0315$.

\section{Comparing VAR, MVAR and DCC}

Dynamic Conditional Correlation models \citep[DCC,][]{Engle2002} are a class
of multivariate GARCH models in which conditional correlations between 
elements of a vector series are time dependent.
In particular, given the conditional covariance matrix of the model at time $t$, $H_t$,
each entry $h_{ij}$ of the matrix is modelled as a univariate GARCH model. 
DCC models are used in finance to predict behavior of vector time series in
which the assets are correlated and heteroskedastic, thanks to the fact that
the conditional covariance matrix of a predictor is always fully specified.

In his model, \citet{Engle2002} states that correlation is based
(i.e. conditional) on information known the previous period, and that correlation
matrices of multi-period
forecasts are similarly defined. According to this, each time point produces a 
different conditional correlation matrix.
A similar argument can be presented for MAR and MVAR models, except we work with 
conditional covariance matrices, rather than correlations. 
The conditional covariance matrix clearly depends on past observations, and one
unique conditional covariance matrix is produced at each time point, except for
some limiting cases. As seen 
in Section~\ref{sec:mvarportfolio}, this also applies to one and multi-step
forecasts. For this reason, we consider a comparison between MVAR and DCC 
to be appropriate.

We compare here the performance of modelling the data in Section~\ref{sec:realdata} with
an MVAR model and a DCC model. We also add a comparison with a fitted
vector autoregressive model of order 3 (VAR(3)). 
We use a rolling-window setup for comparison of the density forecasts.

First,
we consider a window from the first observation on January 2$^{nd}$ 2016 to
October 10$^{th}$ 2019, thus including 766 observations (roughly $90\%$ of the
available data). We use this to estimate $\MVAR$, DCC and VAR models,
and derive one and two steps density forecasts for the three models for
October 11$^{th}$ and October 12$^{th}$, and calculate forecasting accuracy
using their respective observed values. 
We then move the window forward by one day. The window would now contain 
$766$ observations from from the first observation on January 3$^{rd}$ 2016 to
October 11$^{th}$ 2019. 
The procedure is repeated until the window contains the most up to date
observation on January 28$^{th}$ 2020, which allows a one step forecast.
Therefore, we obtain $100$ one step density forecasts, and 
$99$ two step density forecasts.

Comparison of density forecasts is ususally done using scoring rules. Here we
 compare the three models in terms of some strictly proper scoring rules:
  Continuous Ranked Probability Score
  \citep[CRPS, see for instance][]{Gneiting2004crps}, logarithmic score
  \citep[LogS,][]{good1952log}, 
  and the Dawid-Sebastiani score \citep[DSS,]{dawid1999dss}. 
  In all cases, the 
  "preferred" model is the one which minimises the score.

 Given an observation $x$ and the associated forecast distribution $F$, CRPS is
  defined mathematically as:
 \begin{equation}
 CRPS(F, x) = \int_{\mathbbm{R}}\left(F(y) - \mathcal{I}(y\geq x)\right)^2 dy 
 \end{equation}
 where $\mathcal{I}(\cdot)$ is the indicator function assuming value $1$ when the
 argument $y>x$ is true, and $0$ othetwise.
 CRPS is a measure of discrepancy between the forecast CDF, $F$, and the
 empirical CDF of the observation $x$.
Notice that it was purposely chosen not to
involve factor
models in the comparison due to the small number of assets considered.

LogS is calculated as the logarithm of the probability estimate for the observation
with respect to its forecasting distribution:
\begin{equation}
 \text{LogS}\left(
F, x\right)= \log f\left(x\right).
\end{equation}

Given the mean $\mu$ and the variance $\sigma^2$ of the predictive distribution,
DSS is calculated as
\begin{equation}
\text{DSS}\left(F, x\right) = - \log \sigma^2 - \dfrac{1}{\sigma^2}\left(
x - \mu \right)^2
\end{equation}

A DCC-GARCH(1,1) was found to be the best model for the return series
assuming multivariate normal innovations.
For this, and for the VAR(3) model, the same rolling window procedure as for MVAR
is performed. For each forecast of each model, we calculate CRPS, LogS and 
DSS, and take the average score for comparison. Results can be seen in 
Tables \ref{tab:comp1} and \ref{tab:comp2}:

\begin{table}[h]
	\centering
	\begin{tabular}{|c|c|c|c|}
		\hline
		&CRPS  &LogS &DSS\\
	\hline
	$\MVAR(2;3,2,1)$ &$0.004895$ &$-3.300112$ &$-8.410847$ \\
	\hline 
	DCC-GARCH(1,1) &$0.005048$ & $-3.296478$ &$-8.430833$ \\
	\hline
	VAR(3) &$0.005123$ &$-2.930259$  &$-7.698397$\\
	\hline
	\end{tabular}
	\caption{Average scores for one step density forecasts.}
	 \label{tab:comp1}
	\end{table}
\begin{table}[h]
	\centering
	\begin{tabular}{|c|c|c|c|}
		\hline
		&CRPS  &LogS &DSS\\
		\hline
		$\MVAR(2;3,2,1)$ &$0.004805$ &$-3.310742$ &$-8.439584$ \\
		\hline 
		DCC-GARCH(1,1) &$0.004845$ & $-3.326115$ &$-8.490107$ \\
		\hline
		VAR(3) &$0.005022$ &$-3.024682$  &$-7.887240$\\
		\hline
	\end{tabular}
	\caption{Average scores for two step density forecasts.}
	\label{tab:comp2}
\end{table}

|



From this comparison, it appears that the only significant differences between
MVAR and DCC-GARCH, in terms of forecast accuracy, are in the CRPS for the one
step predictor (first column of Table \ref{tab:comp1}), in which on average MVAR
outperforms DCC-GARCH, and for DSS in the two step predictor 
(last column of Table \ref{tab:comp2}), where DCC-GARCH outperforms MVAR instead.
However, neither method is objectively better than the other. On the other hand, 
we notice that the VAR model is far behind in terms of forecasting accuracy,
and therefore may not be suitable for predicting portfolio returns. 
We conclude that our method for portfolio optimisation with
MVAR models may be a valid alternative
to a widely accepted method such as
DCC-GARCH, while it clearly outperforms the standard VAR model.

\section{Conclusions}

The paper presents an innovative way of using mixture vector autoregressive
models for portfolio optimisation. The method consists in deriving analytically
predictive distributions of future observations, and use the conditional covariance
 matrix,
together with modern portfolio theory, to build an efficient portfolio and
obtain a distribution for future returns. We have seen in fact that, assuming
multivariate normal distributions for mixture components, the conditional 
predictive distribution of the portfolio return at a future horizon $h$ 
itself follows a (univariate) mixture of $g^h$ normal components, depending
on observation up to the present.  

The methodology was tested both on a simulated and a real dataset. For the 
latter, we compared performance of MVAR
with the widely used dynamic conditional correlation model, which uses
multivariate GARCH to estimate conditional correlations, and with the 
 VAR model, using a rolling-window forecasting scheme. In particular, 
 forecasting accuracy was assessed using three strictly proper scoring rules,
 averaged over the number of forecasts.
 In terms of minimum variance portfolios, the conclusion was
 that
the MVAR and DCC-GARCH have similar performance on the analysed datasets,
 suggesting
MVAR may be considered a valid alternative to DCC-GARCH. Furthermore, it was
seen that MVAR outperformed VAR.

There are various ways in which the methodology could be
extended.  One possibility is to employ the GMVAR model
\citep{KALLIOVIRTA2016485} in place of the MVAR model. GMVAR has the useful
property that the mixing weights depend on past values of the process.  On the
other hand, the region for the autoregressive parameters of GMVAR is
restricted to a subset of that of MVAR. Also, MVAR and GMVAR have different
dynamics and stationary distributions.  So the two classes of models
complement each other.

Another possible extension of our method is to incorporate factor models, in order
to be able to model a large number of possibly correlated assets at a time.
In addition, distribution assumptions other than normal can be made on the 
innovation terms. For example, considering a distribution with heavy tails 
might need a smaller number of components to fit the data. However, estimation
would become much more complicated, and numerical algorithms would be required.

\appendix
\section{A property of the multivariate normal distribution}

At several places we use the standard result that any linear combination of the
elements of a random vector from the multivariate normal distribution is
univariate normal. More specifically, let $X$ be a random vector of length $m$
following a multivariate normal distribution with mean $\mu$ and covariance
matrix $\Sigma$. Let $a$ be a constant vector of the same length as $X$. Then
\begin{equation} 
a^TX = \sum_{j=1}^{m}a_j X_j \sim 
N(a^T \mu, a^T \Sigma a)
\label{eq:multtouniv}
\end{equation}

\section{Derivation of the representation of $y_{t+2}$ and its characteristic function}
\label{app:derive}

We here derive the analytic expression for the predictor $y_{t+2}$ when information
up to time $t$ is available. We also derive its characteristic function, to show
that the distribution of such predictor is a mixture of $g^2$ normal distributions.
Proof is simply the multivariate version of the proof in \cite{boshnakov2009mar}.
Let $Z_t \in \lbrace 1,\ldots,g$ be the allocation random variable 
defined in Section \ref{sec:MVAR}, and assume
$z_{t+2} = k, z_{t+1} = l$ at times $t+2$ and
 $t+1$. We have that
 
\begin{equation}
\begin{split}
y_{t+2} &= \mu_{t+2, k} + \Omega^{1/2}\varepsilon_{t+2, k} \\
&= \mu_{t+2, k} - \Theta_{k,1}Y_{t+1} + \Theta_{k,1}Y_{t+1} +
 \Omega^{1/2}\varepsilon_{t+2, k} \\
 &= \left(\mu_{t+2, k} - \Theta_{k,1}Y_{t+1} + \Theta_{k,1}\mu_{t+1,l}\right) +
 \Theta_{k,1} \Omega_{l}^{1/2}\varepsilon_{t+1, l} +  \Omega^{1/2}\varepsilon_{t+2, k} \\
 &= \mu_{t+2;k,l} +  \Theta_{k,1} \Omega_{l}^{1/2}\varepsilon_{t+1, l} +  \Omega^{1/2}\varepsilon_{t+2, k}
\end{split}
\end{equation}
where $\varepsilon_{t+h,k}$ is the innovation term associated with
the $k^{th}$ component.

We want an expression that does not contain $Y_{t+1}$. Hence, we rewrite
 $\mu_{t+2;k,l}$ as
\begin{equation*}
  \begin{split}
    \mu_{t+2;k,l} &= \mu_{t+2, k} - \Theta_{k,1}Y_{t+1} + \Theta_{k,1}\mu_{t+1,l}
    \\ &=
    \Theta_{k,0} + \sum_{i=1}^{p} \Theta_{k,i} Y_{t+2-i} - \Theta_{k,1} Y_{t+1} 
    + \Theta_{k,1}\left(\Theta_{l,0} + \sum_{i=1}^{p}  \Theta_{l,i}Y_{t+1-i}\right)
    \\
    &= \Theta_{k,0} + \Theta_{k,1} \Theta_{l,0} - \Theta_{k,1} Y_{t-1} + 
    \Theta_{k,1} Y_{t+1} + \sum_{i=2}^{p} \Theta_{k,i} Y_{t+2-i}
    \\ & \qquad {}
    + \Theta_{k,1} \sum_{i=1}^{p} \Theta_{l,i}Y_{t+1-i}
    \\
    &= \Theta_{k,0} + \Theta_{k,1} \Theta_{l,0} + 
    \sum_{i=1}^{p-1} \Theta_{k, i+1} Y_{t+1-i} +  \Theta_{k,1} \sum_{i=1}^{p} \Theta_{l,i}Y_{t+1-i}
    \\
    &=\Theta_{k,0} + \Theta_{k,1} \Theta_{l,0} + 
    \sum_{i=1}^{p-1} \left(\Theta_{k, i+1} + \Theta_{k,1}\Theta_{l,i}\right)Y_{t+1-i} +
    \Theta_{k,1} \Theta_{l,p} Y_{t+1-p}
  \end{split}
\end{equation*}
And therefore we have the expression for $Y_{t+2}$
\begin{multline*}
    Y_{t+2}
    = \Theta_{k,0} + \Theta_{k,1} \Theta_{l,0} + 
    \sum_{i=1}^{p-1} \left(\Theta_{k, i+1} + \Theta_{k,1}\Theta_{l,i}\right)Y_{t+1-i} +
    \Theta_{k,1} \Theta_{l,p} Y_{t+1-p}
    \\
    {} + \Theta_{k,1} \Omega_{l}^{1/2}\varepsilon_{t+1, l} +  \Omega_k^{1/2}\varepsilon_{t+2, k}
\end{multline*}
We deduce that, given observed $z_{t+2}, z_{t+1}$:
\begin{equation*}
\begin{split}
  &\Ex[Y_{t+2}\mid Z_{t+2}, Z_{t+1}, \mathcal{F}_t]
  \\ & \qquad = \Theta_{k,0} + \Theta_{k,1} \Theta_{l,0} + 
\sum_{i=1}^{p-1} \left(\Theta_{k, i+1} + \Theta_{k,1}\Theta_{l,i}\right)Y_{t+1-i} +
\Theta_{k,1} \Theta_{l,p} Y_{t+1-p} \\
&\Cov\left(Y_{t+2}\mid Z_{t+2}, Z_{t+1}, \mathcal{F}_t\right) = 
\Theta_{k,1} \Omega_l \Theta_{k,1} + \Omega_k
\end{split}
\end{equation*}
We now need to derive the characteristic function for the predictor. Recall the 
characteristic function for the multivariate normal distribution and $Y_{t+1}$ 
can be written as
\begin{equation*}
  \varphi_{t+1 \mid t} \equiv \Ex\left[e^{is^Ty_{t+1}} \mid \mathcal{F}_{t-1} \right]
  = \Ex \left[
          \sum_{k=1}^{g} \pi_k e^{is^T\mu_{t+1;k}} \varphi_k(\Omega_K^{1/2}s)	
        \right]
\end{equation*}
It follows that, for $Y_{t+2}$, we have
\begin{equation*}
\begin{split}
\varphi_{t+2\mid t}(s) & \equiv \Ex\left[e^{is^TY_{t+2}} \mid \mathcal{F}_t\right] =
 \Ex\left[ \Ex \left(e^{is^TY_{t+2}} \mid z_{t+2}, z_{t+1}, \mathcal{F}_t \right) 
 \mid \mathcal{F}_t \right] \\
 &=\Ex\left[is^{T \mu_{t+2;k,l}} \Ex \left(e^{\Theta_{k,1}
 	 \Omega_{l}^{1/2}\varepsilon_{t+1, l} +  \Omega^{1/2}\varepsilon_{t+2, k}} 
  \mid z_{t+2}, z_{t+1}, \mathcal{F}_t \right) \mid \mathcal{F}_t \right] \\
  &= \sum_{k,l=1}^{g} \pi_k \pi_l e^{is^T\mu_{t+2;k,l}} 
  	\varphi_1(\Theta_{k,1}\Omega_l^{1/2} s) 
  	\varphi_2(\Omega_k^{1/2} s)
\end{split}
\end{equation*}
Thus, the conditional distribution of $Y_{t+2}$ given $\mathcal{F}_t$ is 
a mixture of $g^2$ components with mixing weights $\pi_k\pi_l$. For
a normal mixture, we also have that:
\begin{equation*}
\varphi_1(\Theta_{k,1}\Omega_l^{1/2} s) 
\varphi_2(\Omega_k^{1/2} s) = e^{\Theta_{k,1} \Omega_l \Theta_{k,1}^T} 
e^{\Omega_k} = e^{\Theta_{k,1} \Omega_l \Theta_{k,1}^T + \Omega_k}
\end{equation*}
which shows that the conditional distribution of the two-step predictor
is a mixture of Normals with means $\mu_{t+2;k,l}$ and 
covariance matrices $\Theta_{k,1} \Omega_l \Theta_{k,1}^T + \Omega_k$.

\bibliographystyle{kluwer}
\bibliography{mvar}

\end{document}